\documentclass[twocolumn,prb,showpacs,preprintnumbers,amsmath,amssymb,a4paper,floatfix]{revtex4}

\usepackage{graphicx}
\usepackage{dcolumn}
\usepackage{bm}

\begin{document}

\title{An {\it ab initio} effective Hamiltonian for magnetism including
longitudinal spin fluctuations}

\author{S. Shallcross}
 \altaffiliation{Department of Physics and Measurement Technology, \\
University of Link\"oping, SE-581 83 Link\"oping, Sweden.}
\email{sam_shallcross@yahoo.co.uk}

\author{A.~E. Kissavos}
 \affiliation{Department of Physics and Measurement Technology, \\
University of Link\"oping, SE-581 83 Link\"oping, Sweden.}


\author{V. Meded}
 \affiliation{Department of Physics, Uppsala University, \\
Box 530, 751 21 Uppsala, Sweden.}

\author{A.~V.~Ruban}
 \affiliation{Royal Institute of Technology, SE-100 44, Stockholm, Sweden.}

\date{\today}

\begin{abstract}

We discuss the use of the magnetic force theorem (MFT) using different reference states upon which
the perturbative approach is based. Using a disordered local moment (DLM) state one finds good
Curie (or Ne\'el) temperatures, and good energetics for planar spin spirals in the 3d magnets
Fe, Co (fcc), Mn, Cr. On the other hand the ferromagnetic reference state provides excellent energetics 
for small $\theta$ spin spirals in Fe, Co and Ni, and by extension magnon energies under the 
assumption of adiabacity. However, planar spin spiral energetics and transition temperatures for 
Ni, Fe, Mn, and Cr show worse agreement. The reasons for this, and for the case of fcc Co where both approaches 
work very well are discussed.
We further provide an extension of the mapping of the quantum problem to include longitudinal
fluctuations, and discuss the role they will play in magnetic phase transitions. This construction
is tested using planar spin spirals where ${\bf q}$ is fixed but the moment is allowed to relax. It
is demonstrated that results from the magnetic force theorem approach and directly
calculated {\it ab initio} values agree very well.

\end{abstract}

\maketitle

\section{Introduction}

Constructions based upon the mapping of the ground state energetics of a quantum
system onto an appropriate classical model have been found to be of great use in solid
state physics. Direct and accurate quantum mechanical calculations of a system are still only
practically possible for system sizes of the order of hundreds of atoms, and so in order
to explore systems of much larger size the substitution of the quantum problem by
an classical one is neccesary. 
The most prominent example of such a procedure is probably the Ising model in alloy
physics, and a great deal of progress in understanding the ground state and phase behavior 
of binary alloys has been made in this way. In this case the classical variable is
just the site occupation.
In the case of magnetic systems the appropriate classical model is the Heisenberg model,
which in its simplest form reads
\begin{equation}
\label{eq:HM}
E = -\sum_{ij} J^{(2)}_{ij} {\bf s_i}.{\bf s_j}
\end{equation}
where $J^{(2)}_{ij}$ are the exchange parameters and ${\bf s_i}$ a $d$ dimensional 
classical vector.
In the case of magnetism the emergence of such local moment variables is not as
obvious as that of site occupation in the alloy situation. 

In fact the fundamental justification for the
usefulness of such variable comes from a separation of time scales \cite{gyor85}. If one
considers a magnetic material at some low temperature the moment on a particular
site will fluctuate rapidly on a time scale given by the electron hopping. However,
averaged over longer time scales the expectation value of the site moment is a more
stable quantity. In particular, the magnon frequency is often orders of magnitude slower. This
adiabatic condition allows the deployment of a classical Heisenberg model to study the
low temperature magnetic dynamics. Given an accurate procedure for mapping between the quantum
and classical regimes the study of the dynamics of complex materials is facilitated.
A further use, in general less justified, is the 
study of phase transitions in magnetic systems. Here one extends the concept of the local
moment to high temperatures, and assumes that the statistical mechanics of the high 
temperature state is given
simply by the classical partition function of the ground state energy.

There are several ways of performing such a map, which can be divided into
so called structure inverse method and perturbative methods. In structure inverse methods
one extracts the parameters for the Heisenberg model from the total energies 
of a set of magnetic structures. The structures used in this set are most often
spin spirals, and there has been some debate as to whether small $\theta$ spin spiral
or planar spin spiral structures should be used \cite{hal98,rose97}. 
The alternative perturbative approach comes from considering the rotation of a
pair of spins in some reference state by opposite and vanishing angles. This leads
directly to the parameters of a classical Heisenberg model. Note that there is 
no restriction on the form of reference state in this method.

The relative merits of such approaches have been well cataloged in the case of
the analogous alloy situation \cite{ruban04}. The structure
inverse type methods have the advantage that they are in principle limited in accuracy
only by the underlying method used to determine the total energies. On the other
hand the range and importance of specific interactions (which may be more than simply pair
interactions) is not known beforehand, and further it is difficult to systematically converge the
method with respect to the range and type of interactions. A particular problem may arise
if the interactions are further configuration dependent. In perturbative methods
the situation is reversed. With such an approach the convergence of the interaction set
is relatively trivial, and by probing different reference states one can (as will be discussed)
elucidate the configuration dependence if any. On the other hand
such methods make approximations beyond those of the underlying total energy method.

The two original formulations of force theorem based methods were based on the ferromagnetic
\cite{lech84,lech87}
and disordered local moment reference states \cite{ogu83}, we will refer to these
two approaches as ferromagnetic magnetic force theorem (FM-MFT) and disorder local moment
magnetic force theorem (DLM-MFT). The former has become
widely used in the intervening years, but not the latter version. This lack of use
is most likely a combination of the result of the initial calculation of the Curie temperature of Fe,
which turned out to be 2700 K, and the fact that the equilibrium DLM moment in many
systems is substantially different from that of the ground state magnetic structure.

With regard to the first point it was noted recently \cite{ruban04} that the
large value found for the Curie temperature was simply due to numerical error in 
the original work, the correct result being around 1080 K which compares well 
with the experimental Curie temperature of 1040 K.

The second point is more critical, however when one regards the DLM state as
merely a reference medium, and not as a state with which one attempts to describe
the paramagnetic state, then this problem may be circumvented by the use of
a fixed spin moment procedure.

On the other hand, the FM-MFT is also known to suffer from some failures however. Most recently \cite{ruban04} it
was shown that using an accurate underlying method for the electronic structure of the FM
reference state resulted in a Curie temperature of 550 K in the Local Density Approximation,
roughly half the experimental value. A long standing problem has been the Curie temperature of Ni
which has been found to be much smaller than the experimental value of 640 K in all works
using this method.

This paper addresses the question of the role of the effective medium in the mapping of
the quantum problem, and in particular whether certain properties are best calculated with one
or the other approach. In Section 2 we describe briefly the formalism used and give details of
our computations. Section 3 presents results for the energetics of planar and small $\theta$ 
spin spirals, and compares the results of direct calculation with those from the FM-MFT and
DLM-MFT methods. We then present the results of Curie and Ne\'el temperatures calculated with
both methods. In section 5 we introduce an approach for the inclusion of longitudinal fluctuations
in conjunction with perturbative methods, after which we conclude.

\section{Ferromagnetic and Disordered Local Moment exchange integrals}

The magnetic force theorem makes use of Lloyd's formula in
Green's function formalism for the change in
the integrated density of states upon the embedding of some cluster.
For the FM-MFT one arrives at the following equation for the
bi-linear exchange term in the Heisenberg model,

\begin{equation}
\label{eq:J2} 
J^{(2)}_{ij} = \frac{1}{4 \pi} Im \int^{E_F} dz
Tr_L ([P^{\uparrow}_i - P^{\downarrow}_i] g_{ij}^{\uparrow}
[P^{\uparrow}_j - P^{\downarrow}_j] g_{ji}^{\downarrow}
\end{equation}
where $P^{\uparrow}_i$ and $P^{\downarrow}_i$ are the potential functions and $g_{ij}$ elements
of the auxiliary Green's function. The DLM-MFT makes use of the Coherent
Potential Approximation (CPA) for the construction of the disordered local moment 
reference state. Note that the DLM is not here deployed to
model the \emph{paramagnetic} state, which will in general have non-zero short
range order parameters not treatable within CPA, but simply as a reference
medium for the embedding of a spin cluster. The formula for the
exchange parameters becomes now

\begin{equation}
\label{eq:GPM}
J^{(n)}_{ij} =  \frac{1}{4 \pi} Im \int^{E_F} dz
Tr_L (\Delta t_i \tilde{g}_{ij} \Delta t_j \ldots
\Delta t_k \tilde{g}_{ki})
\end{equation}
where $\tilde{g}_{ij}$ are now elements of the auxiliary coherent Green's
function and $\Delta t_i$ is the difference of spin up and spin down on site
scattering $t$ matrices at site $i$ which are related to the potential function 
at site $i$ and coherent potential function by

\begin{equation}
t_i = [1 + (\tilde{P} - P_i) \tilde{g}_0]^{-1} (\tilde{P} - P_i)
\end{equation}

Eq.~\ref{eq:GPM} allows for the calculation of any embedded cluster of spins,
however from the symmetry of the exchange interaction \cite{blanc95} only even clusters
have non zero energy. The full Heisenberg expansion is in fact given by

\begin{eqnarray}
\label{eq:HE}
E & = & -\sum_{ij} J^{(2)}_{ij} {\bf s_i}.{\bf s_j}
    -\sum_{ij} J^{(2-2)}_{ij} ({\bf s_i}.{\bf s_j})^{2}
    - \ldots \nonumber \\
  &   &  -\sum_{ij} J^{(n-n)}_{ij} ({\bf s_i}.{\bf s_j})^{n} \nonumber \\
  &   &  -\frac{1}{4!}\sum_{ijkl} J^{(4)}_{ijkl}
    [({\bf s_i}.{\bf s_j})({\bf s_k}.{\bf s_l}) +
     ({\bf s_j}.{\bf s_k})({\bf s_l}.{\bf s_i}) \nonumber \\
  &   & +    ({\bf s_l}.{\bf s_i})({\bf s_j}.{\bf s_k})]
    - \ldots
\end{eqnarray}

\section{Computational details}

The electronic structure calculations were performed in the Korringa-Kohn-Rostocker (KKR) 
scheme in the atomic sphere approximation (ASA). The basis used contained $s$, $p$, $d$, and $f$ orbitals.
Multipole moment corrections for the charge density up to $l_{max} = 6$ were included.
The exchange correlation functional used was the Local Density Approximation in all cases 
with the Perdew, Burke,
and Ernzerhof parameterization of the results of Ceperly and Alder.
The integration of the Green's function
was taken in the complex plane with 16 energy points on a semi-circular contour.
The experimental room temperature lattice parameters were used in all cases expect
for that of fcc Mn where an expanded lattice parameter was used to ensure a anti-ferromagnetic solution.

\section{Spin spirals}

The energy of a spin spiral in the Heisenberg model is given very simply by a Fourier transform
as

\begin{equation}
\label{eq:sse}
E({\bf q}) = -J({\bf q}) - \sin^2 \theta [J({\bf q}) - J({\bf 0})] - \ldots
\end{equation}
Note that if only bi-linear terms are included in the Heisenberg expansion then
the $\theta$ and ${\bf q}$ dependencies decouple and the difference between a
small $\theta$ spin spiral and a planer spin spiral becomes simply a matter of scale.

One would intuitively expect that, by virtue of its perturbative nature, the 
magnetic force theorem would provide a good description of the magnetisation energy changes for
configurations close to the reference state. Hence in the case of the FM-MFT one must thus
measure energies from the energy of the FM state, and in the case of the DLM-MFT from
the DLM state. One can note that Eq.~\ref{eq:HE} (and hence Eq.~\ref{eq:sse}) measures energies 
relative to the DLM energy, which follows simply from the vanishing of the spin products in this case.

\begin{figure}[floatfix]
\caption{Planar spin spiral energetics for Fe, Co and Ni directly calculated
from KKR-ASA method (filled symbols) and evaluated using magnetic force theorem 
with ferromagnetic (FM, continuous lines) and
disordered local moment reference states (DLM, dashed lines).
}
\vspace{5mm}
\begin{center}
\includegraphics[angle=00,width=0.45\textwidth]{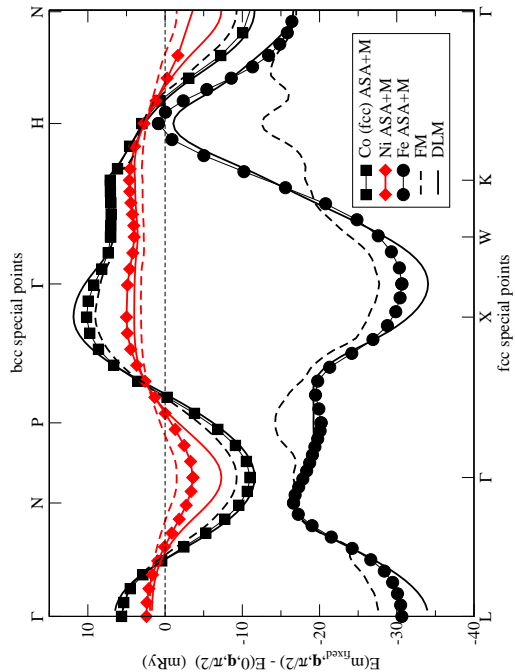}
\end{center}
\label{fig:fm}
\end{figure}

\begin{figure}[floatfix]
\caption{Planar spin spiral energetics for Cr and Mn directly calculated
from KKR-ASA method (filled symbols) and evaluated using magnetic force theorem with ferromagnetic (FM, dashed lines),
antiferromagnetic (AFM, dot-dashed lines), and disordered local moment (DLM, continuous lines) reference states.
}
\vspace{5mm}
\begin{center}
\includegraphics[angle=00,width=0.45\textwidth]{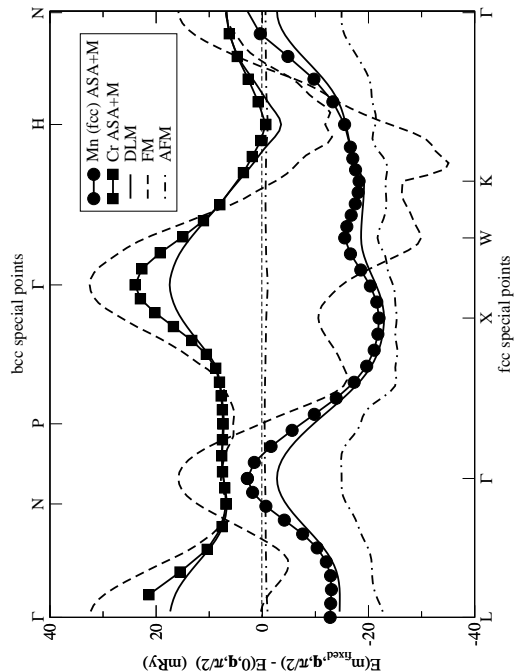}
\end{center}
\label{fig:afm}
\end{figure}

In Fig. \ref{fig:fm} and Fig. \ref{fig:afm} we present the results of direct calculation
of the magnetization energies of planar spin spirals in Ni, Co, Fe, Mn, and Cr with ${\bf q}$ vectors in
the standard reciprocal space paths as indicated. Also shown is the evaluation of Eq.~\ref{eq:sse}
using only bi-linear interactions determined via the FM-MFT and DLM-MFT. For simplicity we show
all energies here measured from the non-magnetic state, which is achieved simply by the addition of
$E_{dlm}$ to Eq.~\ref{eq:sse}. Expect near the $\Gamma$ the perturbation from the FM state 
is significant for planar spin spirals and  indeed one can see
that the agreement is generally better between the DLM-MFT and the directly calculated results,
particularly in the case of the anti-ferromagnets (AFM) Cr and Mn, and Fe near the H point (the special point
generating the AFM structure in real space). In particular, the FM-MFT can hardly be said to work at all
for Cr and Mn. Since the ground state is not FM this is not surprising, and one might expect that
the use of the AFM state as the reference medium would produce better results.
As can be seen in Fig. \ref{fig:afm} this is not the case.

The reason for this failure may come from the neglect of higher order terms in Eq.~\ref{eq:sse}.
For the DLM-MFT we find that the higher order terms in the Heisenberg expansion
are very small in comparison to the bi-linear terms and may be neglected. On the other
hand for the FM-MFT, although we have not directly evaluated the higher order terms, the
"renormalized" bi-linear interactions \cite{ruban04} differ substantially from the bare interactions
which indicates this may be true, in particular for the case of Mn and Fe. In
a weakly ferromagnetic system (in the Stoner sense), such as Fe there will be both spin up and spin
down Fermi surfaces and hence the scattering from spin up to spin down states in Eq.~\ref{eq:J2}
will be greater than in the case of a strong ferromagnet such as fcc Co or Ni.

One can also see that, with the expection of Co,
the DLM-MFT appears to systematically overestimate the stability
of the FM state as compared to the direct calculation. This is particularly pronounced in the case of Ni.
The energetics of these planar spin spirals give an indication of the regime where the DLM-MFT is
applicable in the cases where it fails in the FM limit. It can be seen that for both Ni and Fe the 
deviation between the directly calculated magnetization energy
and the DLM-MFT result is small except for region around the $\Gamma$ point. 
Interestingly, one notes that the FM-MFT actually underestimates the magnetization 
energy of the FM state in all cases.

\begin{figure}[floatfix]
\caption{Small $\theta$ spin spiral energetics for Fe, Co and Ni directly calculated
from KKR-ASA method (filled symbols) and evaluated using magnetic force theorem with ferromagnetic (FM, dashed lines) and
disordered local moment (DLM, continuous lines) reference states.
}
\vspace{5mm}
\begin{center}
\includegraphics[angle=00,width=0.45\textwidth]{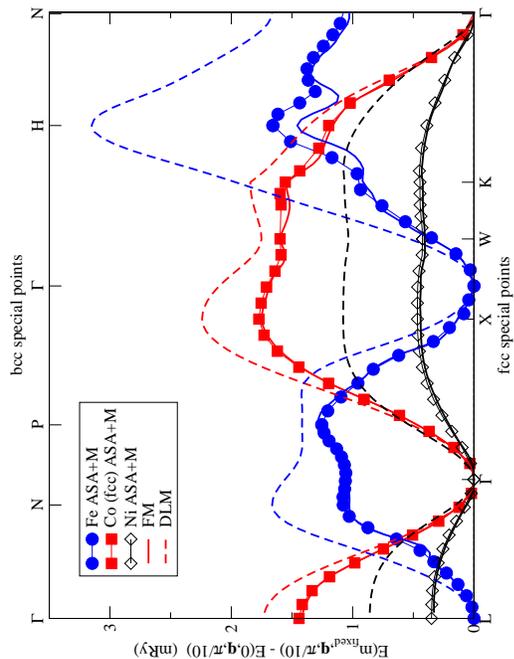}
\end{center}
\label{fig:fmST}
\end{figure}

We now turn to the calculation of small $\theta$ spin spirals. As discussed above it is expected that the
FM-MFT should work very well in this case, and that is indeed seen in Fig. \ref{fig:fmST}. In this
case the energy is
measured relative to the FM energy. In fact, for
Ni and Co the method appears to give practically exact results, whereas for Fe it is slightly off at the
H point. However, this difference might well vanish as $\theta \rightarrow 0$ since the spin spirals
shown here actually had $\theta = \pi/10$, and furthermore the difference with the planar spin spiral
is most pronounced at the H point.
From a linearisation of the Bloch equation of motion the energetics of small $\theta$ spin spirals may be
simply related to magnon energies, and hence the FM-MFT appears to provide 
an excellent way to calculate magnon spectra.

This provides strong computational support for the recent work by Katsnelson {\it et. al~} 
\cite{kat04} who
find the correction to the FM-MFT provided recently by Bruno \cite{bru03} not to be needed for
the calculation of magnon energies, although it might be for Curie temperatures. 
The question as to whether the FM-MFT should be able to provide
correct Curie temperatures, and under what conditions, will be discussed in the next section.

From Fig. \ref{fig:fmST} it is further seen that the DLM-MFT overestimates small $\theta$ spin spirals
by a factor of 2 for Ni and for Fe near the H point. Again, as was seen in the case
of planar spin spirals, for Co the difference between the FM-MFT and the DLM-MFT results is
rather small.

It is interesting to consider the reason for the profound difference between the description of
energetics of spin spirals provided by the FM and DLM magnetic force theorem interactions for the case of
Fe and the anti-ferromagnets Mn and Cr, and on the other hand the good description provided by \emph{both}
approaches for the case of Co. It is well known that hcp and fcc Co are strong ferromagnets
with few minority bands crossing the Fermi energy $E_F$. Since
the effect of non-collinearity is to mix spin up and spin down states, what were previously
orthogonal majority and minority spin bands are no longer so and
thus which could cross in the FM state repel creating hybridisation gaps.
This effect will of course be strongest in weak ferromagnets with a large spin up and spin down density of
states (DOS) at $E_F$, and less important for strong ferromagnets. Thus the reason for the good
description provided by both the FM and DLM reference states in the case Co is the hybridisation is not
important and hence the difference between the FM and non-collinear DLM reference states is less important.
In contradistinction the energetics of Fe, being a weak ferromagnet, should be profoundly affected by this
difference. This also naturally explains why the DLM-MFT gives better agreement in the case of planar 
spin spirals in all cases.

An inspection of the small $\theta$ energetics for Ni shows however that the DLM-MFT does not work very well
in this strong ferromagnet. The reason for this is most likely the itinerant nature of the magnetism in
Ni. 

\section{Curie temperature and spin stiffness calculations}

\begin{figure}[floatfix]
\caption{Relaxation of the magnetic moment of planar spin spirals in Ni, Co, Fe, Mn, and Cr.
Shown are both the results of direct calculation via KKR-ASA method (open circles) and evaluation from
the m-dependent magnetic force theorem approach.}
\vspace{5mm}
\begin{center}
\includegraphics[angle=00,width=0.45\textwidth]{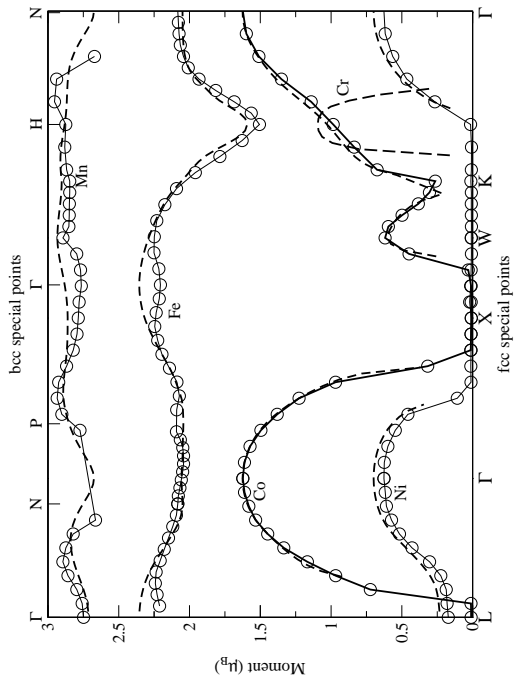}
\end{center}
\label{fig:mom}
\end{figure}

\begin{figure}[floatfix]
\caption{Relaxation energy of planar spin spirals in Ni, Co, and Fe.
Shown are both the results of direct calculation via KKR-ASA method (open and filled symbols) and evaluation from
the m-dependent magnetic force theorem approach (dashed lines).}
\vspace{5mm}
\begin{center}
\includegraphics[angle=00,width=0.45\textwidth]{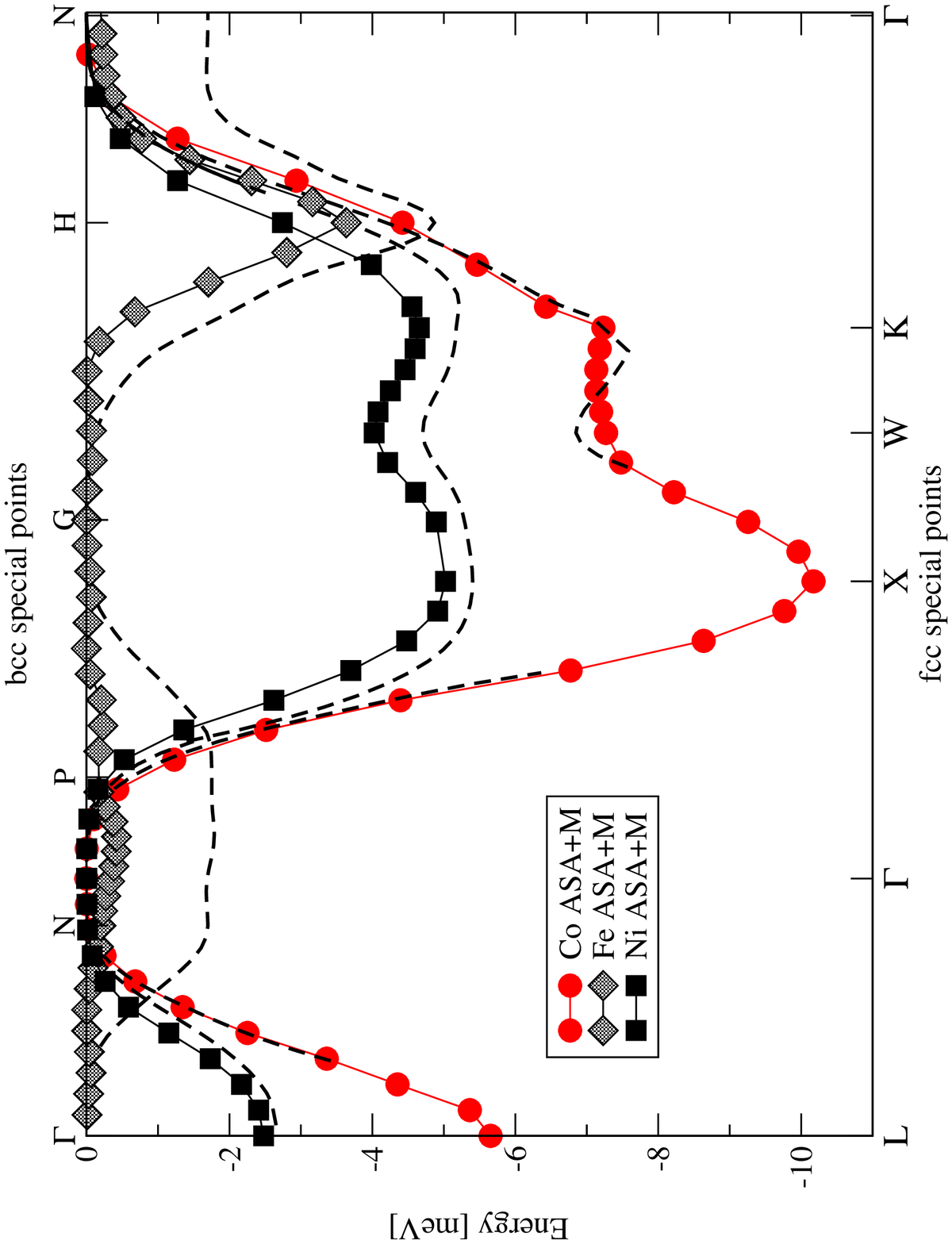}
\end{center}
\label{fig:RE}
\end{figure}

Using the interactions derived via either the FM or DLM versions of the MFT one may perform
Monte Carlo calculations to determine the magnetic transition temperature. Results of 
this procedure are shown in Table \ref{tab:fxprop}.

The arguments given above explain the result for Fe found in Ref. \onlinecite{ruban04}, however
they also allow for the following criteria for the FM-MFT to provide accurate Curie temperatures to be
formulated: The FM-MFT should work for strong local ferromagnets, where it will agree with DLM-MFT, and
not otherwise. The results of Table \ref{tab:fxprop} show that this is indeed the case, with the
Curie temperatures for Co being close in both approaches but there existing a large difference
for Ni and Fe. It is extremely interesting that the Curie temperature for Ni comes out to
be 820 K which is \emph{higher} than the experiment. The result from FM-MFT is 320 K which agrees with 
previous works.

The reason for the overestimate of the Curie temperature comes back
to the imposition of a fixed spin moment constraint on the DLM state. This is neccesary for
the use of the DLM as a reference medium for the force theorem argument, but is artificial
for Ni near the Curie temperature. There the local moment will certainly be lower than
the ground state moment and hence it is that which should be used, and this will have the effect
of \emph{lowering} the Curie temperature from that obtained with the ground state moment. 
Thus if the moment was allowed to take on its true value, the agreement between experiment
and theory may be quite reasonable for Ni.
Of course, the DLM
state has zero equilibrium moment for Ni and so it can not simply be allowed to be
a free parameter, as works well for Fe \cite{ruban04}. A way in which progress can be 
made is described in
the next section.

Near the $\Gamma$ point the magnon spectra behaves as $D q^2$ and this fact is easily used to 
show that $D$ can be expressed in terms of the interactions as
\begin{equation}
\label{eq:D}
D = \frac{2\mu_B}{3 m} \sum_j J_{0j} {\bf R}_{0j}^2
\end{equation}
where $m$ is the magnetic moment of the ferromagnetic state. Results for the spin stiffness
of the ferromagnetic 3d metals are shown in Table \ref{tab:fxprop}. The FM-MFT
provides values in reasonable agreement with experiment, as expected, however the
DLM-MFT also does for Co and Fe. For Co this may be expected from the arguments above, however
for Fe it is more surprising since there appear pronounced differences between the interactions
in the FM and DLM states.
Due to the long ranged nature of the interactions derived from the FM-MFT, D was evaluated
from the derivative of the magnon spectra at the $\Gamma$ point directly. For the interactions
derived from the DLM-MFT Eq.~\ref{eq:D} may be used due to their much quicker decay in real space.

\begin{table*}
\caption{\label{tab:fxprop} Transition temperatures, spin stiffness, and $j_0$ for 3d magnets.
}
\begin{ruledtabular}
\begin{tabular}{lccccccccc}
           & \multicolumn{3}{c}{T$_c$} & \multicolumn{3}{c}{Spin stiffness} & \multicolumn{3}{c}{$j_0$} \\
 & DLM & FM & Exp. & DLM & FM & Exp. & DLM & FM & KKR-ASA \\ \hline
Ni         & 820  & 320  & 624-631   & 1796 & 541 & 555,420 & 3.71  & 9.50  & 5.60  \\
Co (fcc)   & 1350 & 1120 & 1388-1393 & 520  & 480 & 580,510 & 15.51 & 15.50 & 15.09 \\
Fe         & 1190 & 550  & 1044-1045 & 313  & 322 & 280,330 & 15.58 & 9.26  & 12.13 \\
Mn         & 450  & -    & -         & -    & -   & -       & -     & -     & -     \\
Cr         & 421  & -    & 321       &      &     &         & -     & -     & -     \\
\end{tabular}
\end{ruledtabular}
\end{table*}

\section{Inclusion of longitudinal fluctuations}

\begin{table}
\caption{\label{tab:fprop} Magnetic moments calculated directly from KKR-ASA and from magnetic
force theorem.
}
\begin{ruledtabular}
\begin{tabular}{lcc}
         & $\mu_{DLM}$  & $\mu_{KKR}$ \\ \hline
Ni       & 0.69       & 0.62      \\
Co (fcc) & 1.62       & 1.62      \\
Fe       & 2.35       & 2.22      \\
Mn       & 2.86       & 2.79      \\
Cr       & 1.09       & 0.87      \\
\end{tabular}
\end{ruledtabular}
\end{table}

The essential reason for the overestimate of the Curie temperature was that the classical Heisenberg
Hamiltonian allows for only \emph{transverse} fluctuations. This means there is no way for the
size of the Ni moments to respond to the energy cost of orientational disorder by reducing
the exchange splitting. A number of authors have proposed ways to lift this constraint \cite{uhl95,rose97}.
In both of these methods the parameters of the classical model were obtained by the use of the
structure inverse method. What both methods have in common is the addition of an on-site term
to Eq.~\ref{eq:HM} which for our purposes can be written as

\begin{equation}
\label{OS}
E = \sum_{i} J^{(1)}_{i}(m_i^2)
-\sum_{ij} J^{(2)}_{ij}(m_i,m_j) {\bf s_i}.{\bf s_j}
\end{equation}
where m is the magnetic moment. The question now is how to determine $J^{(1)}_{i}$ in
a way consistent with the MFT. An obvious way that suggests itself is a modified application
of the structure inverse method whereby one demands that the DLM energy is well reproduced.
Since under complete spin disorder all spin products vanish Eq.~\ref{OS} reduces to $E_{dlm} = J^{(1)}$,
where for simplicity of expression we consider only simple lattices having one inequivalent lattice. The
$m$ dependence of both $E_{dlm}$ and the exchange integrals $J^{(2)}_{ij}$ 
may then be easily obtained analytically by fitting polynomials to a set of such
values calculated for different $m$.
This forms a very natural link with the DLM-MFT where the DLM reference medium is used itself for
the perturbation theory that results in the exchange parameters, however there is no reason
why it cannot be used with the FM-MFT as well. In this section we shall consider only the former
version though.

The effectiveness of this theory may be tested by again calculating planar spin spirals but 
with the magnetic moment allowed to take on the equilibium value, the $m$-dependent DLM-MFT (referred to
here on in as mDLM-MFT) should then reproduce the {\it ab initio} equilibrium moments and relaxation energies.
It is interesting to see initially how the theory works in the FM limit and these results are
presented in Table \ref{tab:fprop}. As expected the Co moment is well reproduced, but more surprisingly
there is a reasonable agreement in all other cases too. The worst cases, as would be expected, are
Fe and Cr.

In Fig. \ref{fig:mom} are shown the equilibrium moments of the same planar spin spirals calculated with
fixed spin constraint in Section 3. As can be seen the agreement between {\it ab initio} results
and the mDLM-MFT approach is quite reasonable in all cases. In Fig. \ref{fig:RE} is shown the relaxation 
energy of the spin spirals, defined simply as the difference of the energy between the spin spiral with
its equilibrium moment and the spin spiral with the moment fixed to that of the equilibrium ground state 
structure (either FM or AFM).

\section{Conclusions}

We have presented a study of the effect of the choice of the effective medium in approaches
based on the magnetic force theorem. We find that for weak or itinerant ferromagnets it is
essential to use the DLM-FM reference state, but that for strong local moment ferromagnets both
approaches may be used interchangeably. We have further proposed a method whereby
\emph{longitudinal} and well as transverse fluctuations may be incorporated into methods
based on the magnetic force theorem.

\begin{acknowledgments}

Support from the Swedish Research Council(VR) and the Swedish Foundation for Strategic
Research(SSF) is acknowledged.

\end{acknowledgments}


\end{document}